\documentclass[conference]{IEEEtran}
\IEEEoverridecommandlockouts
\usepackage{cite}
\usepackage{amsmath,amssymb,amsfonts}
\usepackage{algorithm}
\usepackage{algorithmic}
\usepackage{graphicx}
\usepackage{textcomp}
\usepackage{xcolor}

\usepackage{bm}
\usepackage{subcaption}
\usepackage{graphicx}
\usepackage{enumitem}
\usepackage{bbold}
\usepackage{braket}
\usepackage{float}
\usepackage{algorithmic}
\usepackage{fancyhdr}
\usepackage{qcircuit}
\usepackage[hyphens]{url}

\def\BibTeX{{\rm B\kern-.05em{\sc i\kern-.025em b}\kern-.08em
    T\kern-.1667em\lower.7ex\hbox{E}\kern-.125emX}}
\begin{document}

\title{Adaptive Circuit Learning for Quantum Metrology\\
\thanks{This work is funded in part by EPiQC, an NSF Expedition in Computing, under grants CCF-1730082/1730449; in part by STAQ under grant NSF Phy-1818914; in part by DOE grants DE-SC0020289 and DE-SC0020331; in part by NSF OMA-2016136 and the Q-NEXT DOE NQI Center; in part by NSF grant OMA-1936118, NSF OAI-2040520, SNSF PP00P2\_176875, and the Packard Foundation (2013-39273).
}}

\author{
\IEEEauthorblockN{ Ziqi Ma\IEEEauthorrefmark{1}, Pranav Gokhale, Tian-Xing Zheng, Sisi Zhou, Xiaofei Yu, Liang Jiang, Peter Maurer, Frederic T. Chong\IEEEauthorrefmark{2}}
\IEEEauthorblockA{University of Chicago, Chicago, IL, 60637, USA}
}

\maketitle

\begin{abstract}
Quantum sensing is an important application of emerging quantum technologies. We explore whether a hybrid system of quantum sensors and quantum circuits can surpass the classical limit of sensing. In particular, we use optimization techniques to search for encoder and decoder circuits that scalably improve sensitivity under given application and noise characteristics.

Our approach uses a variational algorithm that can learn a quantum sensing circuit based on platform-specific control capacity, noise, and signal distribution. The quantum circuit is composed of an encoder which prepares the optimal sensing state and a decoder which gives an output distribution containing information of the signal. We optimize the full circuit to maximize the Signal-to-Noise Ratio (SNR). Furthermore, this learning algorithm can be run on real hardware scalably by using the ``parameter-shift'' rule which enables gradient evaluation on noisy quantum circuits, avoiding the exponential cost of quantum system simulation. We demonstrate up to 13.12x SNR improvement over existing fixed protocol (GHZ), and 3.19x Classical Fisher Information (CFI) improvement over the classical limit on 15 qubits using IBM quantum computer. More notably, our algorithm overcomes the decreasing performance of existing entanglement-based protocols with increased system sizes.
\end{abstract}


\begingroup\renewcommand\thefootnote{\IEEEauthorrefmark{1}}
\footnotetext{Ziqi Ma is now at Microsoft, this work was done at University of Chicago}
\begingroup\renewcommand\thefootnote{\IEEEauthorrefmark{2}}
\footnotetext{Disclosure: Fred Chong is Chief Scientist at Super.tech and an advisor to Quantum Circuits, Inc.}

\section{Introduction}
In recent years, we have seen enormous growth in emerging quantum technologies that exploit quantum mechanics for various applications, such as computation, communication, and sensing. The sensitivity of quantum states to changes in the external environment, while seen as an obstacle in computation and communication, becomes a valuable advantage in sensing. Quantum sensing is believed to have the most immediate real-world impacts, likely before other quantum technologies \cite{pezze2018quantum}. The applications span a wide range of areas, including timekeeping \cite{ludlow2015optical}, spectroscopy \cite{leibfried2004toward}, tests of fundamental physics \cite{dixit2018detecting}, and probing nanoscale systems such as condensed matter and biological systems \cite{abobeih2019atomic}. There has been exciting experimental progress on various physical platforms, such as atomic vapor, trapped ions, Rydberg atoms, superconducting circuits, and nitrogen-vacancy centers in diamond. Even today, practical quantum sensors such as SQUID magnetometers \cite{farhi2014quantum}, atomic vapors, and atomic clocks \cite{ludlow2015optical} have already become the state-of-the-art in magnetometry and timekeeping \cite{degen2017quantum}. 

\begin{figure}
\centering
\includegraphics[width=0.6\columnwidth]{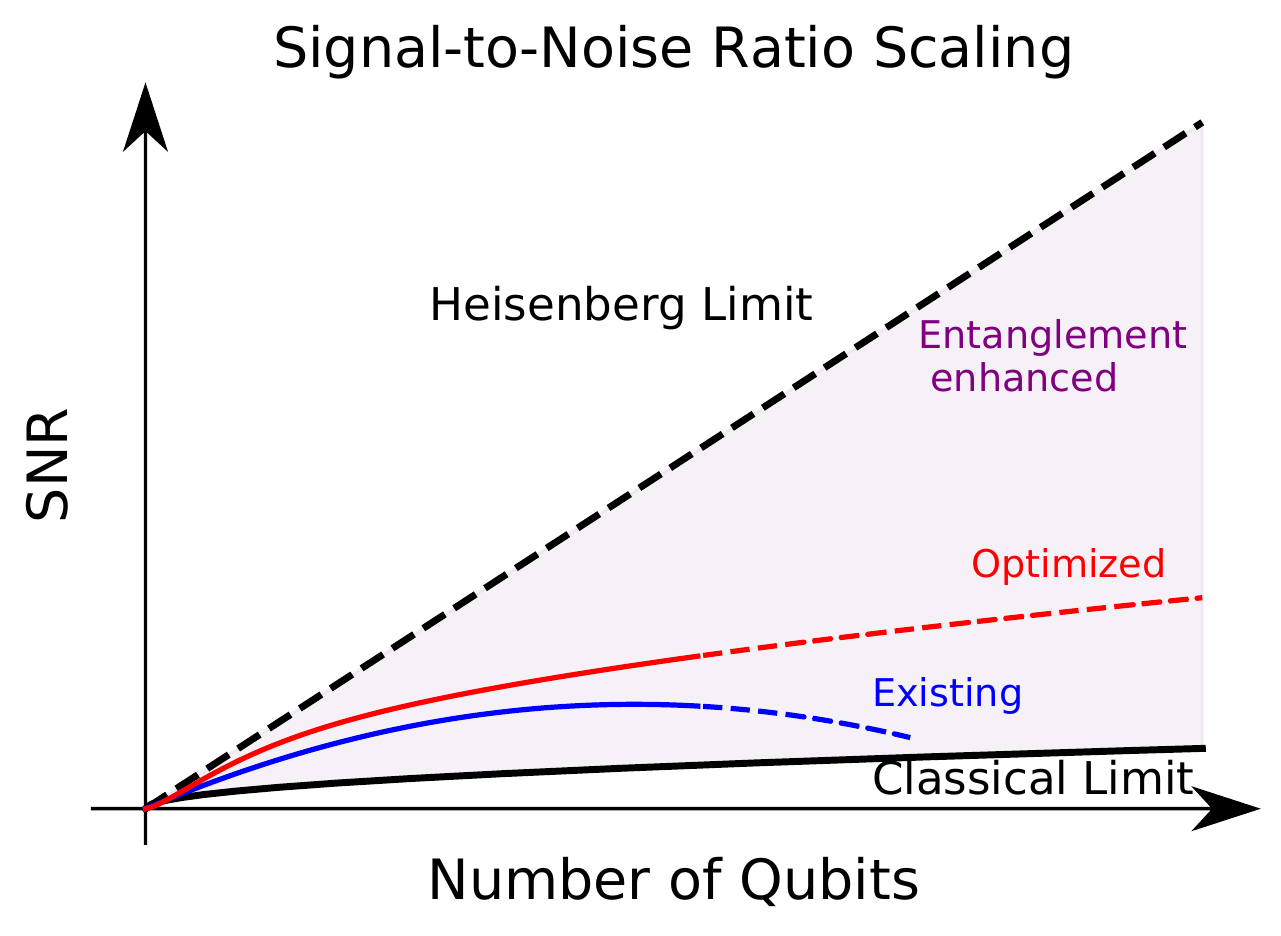}
\caption{
SNR scaling. The classical parallel scheme (Classical Limit) gives square root scaling of SNR versus the number of sensing qubits. The theoretical upper bound of quantum sensing is the Heisenberg Limit, which gives linear scaling. Shaded region corresponds to entanglement-enhanced sensing. Existing and optimized curves are based on experiments on IBM hardware (dotted lines are extrapolations)---optimization is able to overcome the decreasing performance of existing entanglement-based protocols.
}
\label{fig:scaling}
\end{figure}

\begin{figure*}[h]
\centering
\includegraphics[width=1.2\columnwidth]{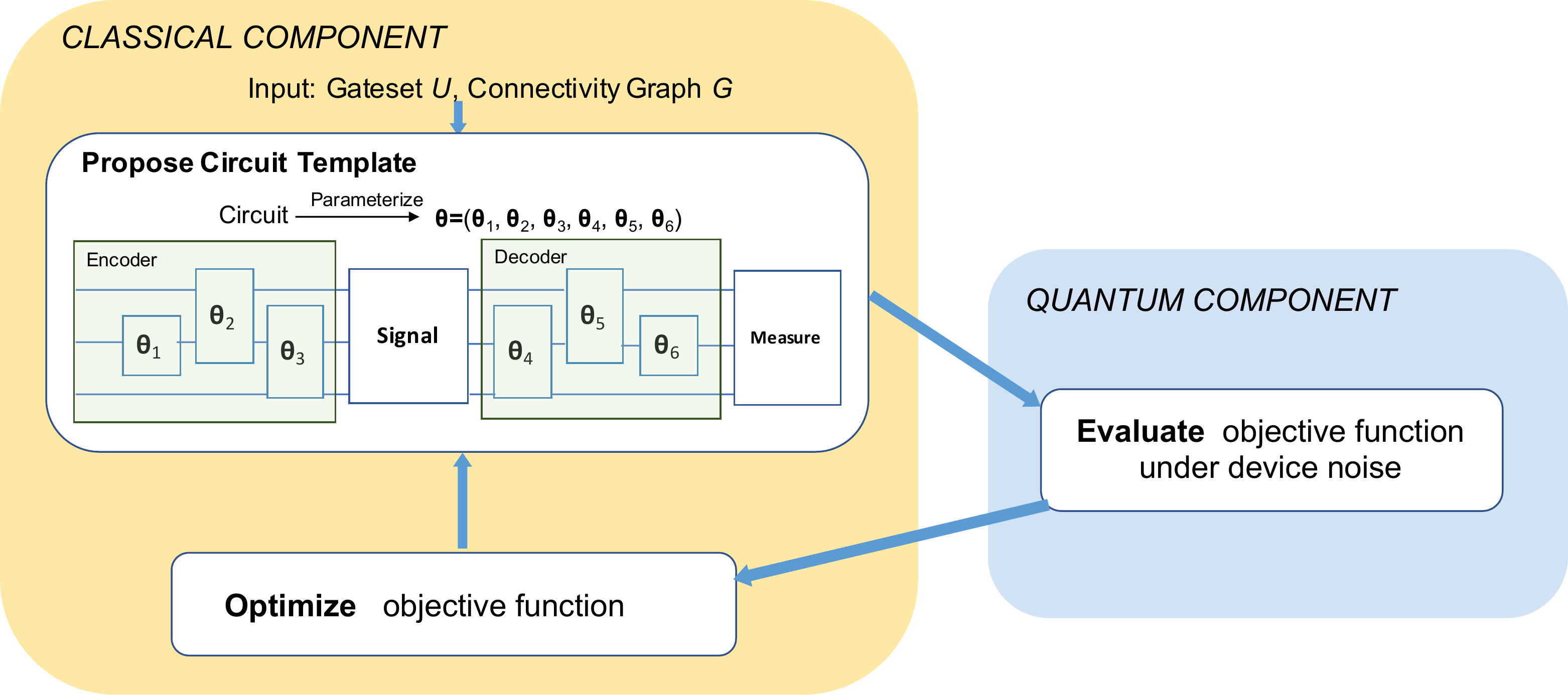}
\caption{Algorithm schematic. Given available gateset and connectivity graph, multiple ansatz's are proposed. Each template parameterizes a circuit into a continuous vector, upon which we run optimization to maximize sensitivity. Optimization could be run on-device via the ``parameter-shift'' rule for gradient evaluation. Optimization across different circuit structures converges upon a final optimized output.}
\label{fig:schematic}
\end{figure*}

Quantum advantage is enabled by entanglement, which allows for higher sensitivity than what can be achieved by a classical parallelization of the sensing qubits---the classical limit. As shown in Figure~\ref{fig:scaling}, the classical limit gives a square root scaling of Signal-to-Noise Ratio (SNR) versus the number of sensing qubits. By contrast, the theoretical limit---the \textbf{Heisenberg limit}---achieves linear scaling \cite{pezze2018quantum,giovannetti2004quantum,giovannetti2011advances,toth2014quantum}. In recent years, there has been exciting experimental progress beyond the classical limit in areas ranging from spectroscopy \cite{leibfried2004toward} to precision measurement \cite{schleier2010states,appel2009mesoscopic} to the famous LIGO experiment for gravitational wave detection \cite{aasi2013enhanced,yu2020quantum}. While the state-of-the-art shows great promise in achieving quantum advantage in sensing, we still expect further improvement---the Hilbert space of entangled states is large and existing protocols only explore a small set of sensing states. Furthermore, the current state-of-the-art in entanglement-enhanced sensing is still mostly proof-of-concept experiments, and such protocols do not necessarily yield good performance on real hardware---each platform is subject to a unique set of noise and control constraints. Given the constraints of limited and imperfect control, device-specific noise, and readout errors of practical hardware, theoretically-optimal sensing protocols yield suboptimal performance. For example, the Greenberger–Horne–Zeilinger state (GHZ state) is optimal without noise but decoheres easily in noisy cases. Another theoretically optimal state, called the spin-squeezed state, is hard to create if we do not have global interaction or all-to-all qubit connectivity \cite{degen2017quantum}. We believe that a co-design optimization approach, which optimizes the sensing apparatus circuit based on characteristics of the underlying hardware, noise model, and signal can provide a solution.

We propose a flexible, architecture-aware circuit structure design, as well as a hardware-based optimization procedure that could be run under device noise.

The algorithm schematic is shown in Figure~\ref{fig:schematic}. We divide the sensing circuit into four components: encoder, signal accumulation, decoder, and measurement. Given the available control gateset, platform-specific noise, and gate/readout errors, our approach aims to find the optimal encoder/decoder pair. The encoder prepares the sensing state prior to exposure to the signal by applying superposition and entanglement to the initial state. Then the sensing qubit(s) will be exposed to the signal for a certain amount of time, which results in changes to the state due to the signal as well as interrogation noise (noise during signal accumulation). A long probing time magnifies the signal but also brings in more noise. The decoder's function is to transform the post-signal state into some other state such that measurement of the qubits yields the greatest information about the signal. Prior work \cite{koczor2019variational,meyer2020variational,PhysRevLett.118.150503, yang2020probe} focuses much more on the encoder for preparing a sensing state, and not so much on the decoder for extracting quantum information into classical information. However, as we show in this paper, given imperfect control and readout errors, decoder design has considerable effects on sensitivity.

Our optimization approach is as follows: to constrain the exponentially large Hilbert space of all possible entangled states, we propose ansatz's based on both physics principles and hardware capacities. The ansatz parameterizes a circuit into a real-valued vector, framing circuit design into a multivariate optimization problem. We use an objective function based on estimation theory, as well as prior results on noisy gradient estimation \cite{schuld2019quantum}, which allows for efficient and scalable search evaluated on real hardware.

We achieve \textbf{3.19x CFI improvement} over classical limit, and \textbf{13.12x SNR improvement} over GHZ protocol on 15 qubits using an IBM quantum computer, with larger gains expected for increased system size. More importantly, we are able to \textbf{overcome the decreasing performance of known entanglement-based protocols, obtaining consistent sensitivity gain from additional qubits}. Compared to the classical limit, our optimization result is equivalent to 3.19x savings in the number of qubits needed, or 3.19x savings in total sensing time to achieve the same SNR. Considering the increasing difficulty of engineering large quantum systems, a 3.19x reduction in system size has significant practical benefits. Furthermore, a smaller system size also allows for smaller sensing volume, which is key in invasive applications such as biological sensing. Reduction of the sensing time, in addition, allows for improved precision in sensing any time-dependent signal.

We enable quantum sensing to be implemented with high fidelity on real near-term noise-prone quantum hardware. Our solution considerably extends the state-of-the-art by incorporating noise-awareness, realistic classical-quantum interfacing, and adaptivity to diverse physical platforms and applications. The main contributions of this work are as follows:
\begin{itemize}
    \item Enabling a guided exploration of a larger space of entangled states than existing protocols;
    \item Leveraging platform-specific information via on-device training under realistic hardware and noise constraints;
    \item Optimizing full sensing circuit including information extraction (rather than just state preparation), which we show to have important effects on performance, especially given gate noise and readout errors;
    \item Considering a holistic noise model including gate noise, interrogation noise, and readout noise, and demonstrating automatic adaptation to different relative magnitudes of these noises;
    \item Exploiting non-uniform signal distributions to further improve sensitivity;
    \item Designing the algorithm to be scalable to large systems and generalizable to different sensing platforms and applications---from nanoscale field sensing to timekeeping to testing physics beyond the standard model.
\end{itemize}

The rest of the paper is organized as follows: Section~\ref{sec:prior} compares this paper to prior work in quantum sensing, quantum-classical hybrid algorithms, and NISQ quantum computing architecture. Section~\ref{sec:background} covers the background of quantum sensing including the theoretical framework, as well as introduces baseline protocols. Section~\ref{sec:optimization} describes our optimization algorithm in detail, discussing the tradeoffs we need to balance as well as our design decisions. Section~\ref{sec:expresults} shows experimental results on IBM hardware, and Section~\ref{sec:simresults} shows simulation results under different noise combinations and with different signal distributions. Section~\ref{sec:outlook} discusses various sensing platforms and applications on which our methodology can be applied, as well as future directions.

\section{Comparison to Prior Work}
\label{sec:prior}
Our work builds on prior work both in three areas: classical-hybrid quantum sensing \cite{koczor2019variational, kaubruegger2019variational, PhysRevLett.118.150503, meyer2020variational}, variational algorithms, \cite{kandala2017hardware,farhi2014quantum,anschuetz2019variational}, and NISQ quantum computing architecture \cite{tannu19asplos,murali19asplos, murali19isca, poulami19micro, murali20isca}.From the sensing perspective, existing classical-hybrid sensing protocols are all evaluated in simulation, consider only limited noise sources, use fixed ansatz structures (thus cannot generalize easily to different platforms when additional constraints are necessary e.g. no individual addressibility on NV platform). They also require deep circuits or global interactions that are impractical on many platforms. Finally, existing variational sensing work focuses on the "encoding" rather than “decoding” part, which we show to be important.

From the variational algorithm perspective, we borrow the idea of alternating between execution on quantum hardware and execution on a classical optimizer which guides subsequent iterations of quantum operations. By limiting the number of operations on quantum hardware per iteration, variational methods reduce errors while maintaining the quantum advantage. Variational algorithms also allows for guided exploration of the large Hilbert space for entanglement generation.

From the architecture perspective, quantum sensing is a new architecture application on real emerging systems, yet the goal for sensing circuit optimization is very different from QC circuit optimization. For sensing we optimize the \textbf{protocol} rather than the \textbf{implementation} of a fixed logical circuit (as in QC), and can leverage noisy operations (if the benefit from entanglement outweighs noise) rather than only selecting a subset of best qubits/gates. Furthermore, we co-optimize the signal accumulation time, the result of which could be much longer than normal QC gates (depending on the platform), meaning sensing circuit could operate in a much higher-decoherence regime compared to QC, and have much higher noise tolerance (as long as we are still able to decode). Most existing circuit-level compilation methods cannot directly apply to sensing, but we see this design space as an exciting area of future architecture work (in aspects such as scalable and noise-aware ansatz design, robust optimization, pulse-level optimization with modified objectives), similar to the recent innovations in QC architecture.

\section{Background}
\label{sec:background}

%For the simulations and experiments in this paper, we mainly consider magnetometry-based sensing, where the signal is a $R_z$ rotation, the angle of which is the product of a frequency value dependent on the magnetic field and probing time $t$. The encoder and decoder in our circuit are sequences of parameterized single-qubit or entanglement gates.

\subsection{Theoretical Framework: Classical and Quantum Fisher Information}
\label{sec:CFI}
Given a specific signal $\omega$, the objective of quantum sensing is to maximize the Signal-to-Noise Ratio (SNR), $\frac{\omega}{\sigma_{\hat{\omega}}}$. SNR, however, is not a good metric for a sensing circuit since it is signal-dependent. If we view the sensing problem as parameter estimation, i.e. constructing an estimator of the unknown signal based on measured outcomes, we could use concepts from estimation theory as better metrics. Based on estimation theory, Classical Fisher Information (CFI) and Quantum Fisher Information (QFI) are used as signal-independent metrics that quantify information \cite{braunstein1994statistical,fisher1925theory}. In particular, QFI quantifies information carried by the quantum state whereas CFI quantifies information in the classical distribution obtained after repeated quantum measurements.

Because noise causes attenuation of signal information, QFI decreases monotonically after each noisy operation. A noiseless measurement in an optimal basis can extract full quantum information into classical information, in which case the CFI is equal to QFI of the final state. However, when measurement cannot be performed on an arbitrary basis or is noisy, the quantum information cannot be fully extracted, and thus CFI is lower than QFI. We find this to be the case in practical sensing applications. Therefore, while prior works such as \cite{koczor2019variational} focus on QFI, we adopt CFI as a more practical metric.

Mathematically, CFI provides a lower bound on the variance of signal estimator $\sigma_{\hat{\omega}}$ via the Cram\'er-Rao Bound \cite{cramer1999mathematical}, therefore upper-bounding SNR. This bound tells us how good the SNR could be given perfect post-processing. With $M$ repeated experiments, the full relation is written as:
\begin{equation}
\sigma_{\hat{\omega}} \geq \frac{1}{\sqrt{M \times \tilde{\text{CFI}}(\omega)}} \geq \frac{1}{\sqrt{M \times \text{QFI}(\omega)}}
\label{eqn:cfi_bound}
\end{equation}
where CFI is defined as
\begin{equation}
    \tilde{\text{CFI}}(\omega)=\mathbb{E}\left[(\frac{d\log(\text{Pr}(X\vert \omega))}{d\omega})^2\right]
\label{eqn:cfi_defi}
\end{equation}
where X is the measurement outcome with a distribution over $2^N$ possible values for a $N$-qubit circuit. Whereas we find the CFI-QFI bound hard to saturate on practical hardware, we find it easy to saturate the first bound via a local 2-degree polynomial fit. Since we are only estimating one signal parameter, this bound is saturated by the Maximum-Likelihood Estimator. Therefore, in this work, we study QFI to understand the signal attenuation at each operation, but use CFI for the final optimization objective because it's more practical.

Note that the definition of CFI in Equation~\ref{eqn:cfi_defi} requires gradient evaluation, which is challenging on noisy hardware. This could be resolved by using the ``parameter-shift'' rule detailed in Section~\ref{sec:parameter_shift}.

\subsection{Parameter-Shift Rule for CFI Evaluation}
\label{sec:parameter_shift}
As shown in Section~\ref{sec:CFI}, CFI evaluation requires gradient evaluation. Note that even if we don't use a gradient-based optimizer, evaluation of our optimization objective still requires gradient. The ``parameter-shift'' rule makes such evaluations possible on noisy hardware. When running on real hardware, we do not have the analytic form of output probabilities and only have access to the noisy sensing circuit. Numerical derivatives via finite differences fail to work since the shot noise and machine noise result in larger fluctuation than a slight change in the signal. A method called the ``parameter-shift'' rule, similar to backpropagation in neural networks, resolves this problem by enabling analytic gradient evaluation \cite{schuld2019quantum, meyer2020variational, PhysRevLett.118.150503}. As shown in Figure~\ref{fig:parametershift}, the gradient at angle value $\theta$ can be estimated with function values at $\theta + \frac{\pi}{2}$ and $\theta - \frac{\pi}{2}$, and this is proven to be true even with noise \cite{meyer2020variational, PhysRevLett.118.150503}. As detailed in Appendix A of \cite{meyer2020variational}, noises such as dephasing and depolarizing channels satisfy this assumption. Mathematically, in our application, gradient of probability of measuring n on signal $\omega$ could be written as $\frac{d\text{Pr}(n|\omega)}{d\omega} = \frac{t^2}{2}(\text{Pr}(n\vert\omega+\frac{\pi}{2}) - \text{Pr}(n\vert\omega-\frac{\pi}{2}))$. Note that this formula is for the single-qubit case. For an $N$-qubit sensing circuit, we use a simple extension with 2$N$ additional circuit evaluations, each time adding/subtracting $\frac{\pi}{2}$ to one qubit and keeping other qubits constant. The ``parameter-shift'' rule allows for macroscopic step sizes instead of microscopic step sizes and is therefore robust to noise.

\begin{figure}[h]
\centering
\includegraphics[width=0.7\columnwidth]{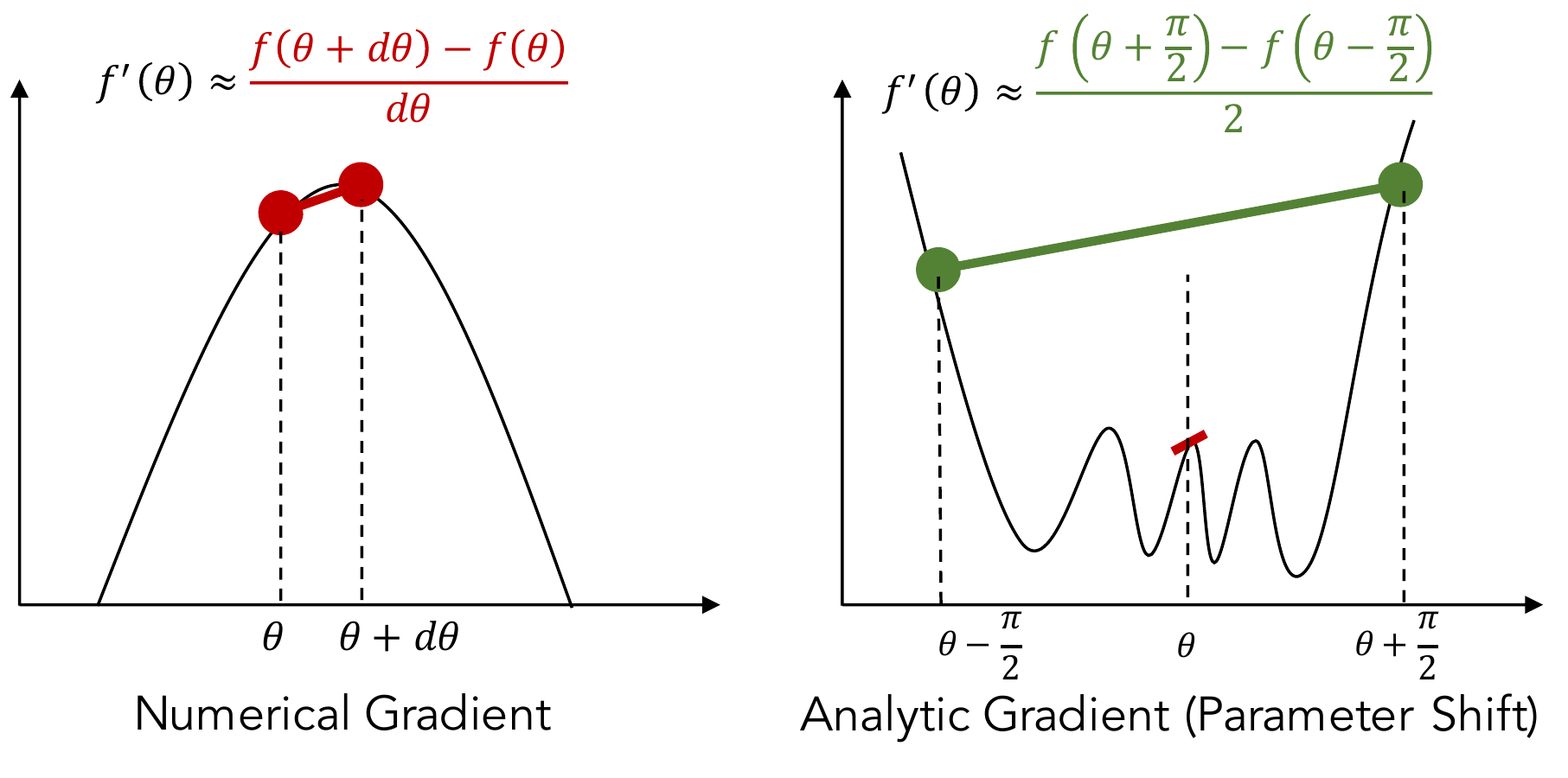}
\caption{Parameter-shift rule for gradient estimation on noisy circuits: to estimate the derivative of a quantum observable, the naive finite-difference approximation fails on practical hardware---true derivative is overshadowed by noise. The ``parameter-shift'' rule proves that gradient at $\theta$ could be estimated with function values at $\theta + \frac{\pi}{2}$ and $\theta - \frac{\pi}{2}$, which resolves this problem. Our CFI evaluation requires gradient of the signal, which is a parameter used N times (on N qubits). This requires a simple extension of the parameter-shift rule: instead of two additional evaluations, we need 2N additional evaluations, each time adding/subtracting $\frac{\pi}{2}$ to one out of the N qubits. The "parameter-shift" rule gives numerically-stable gradient values, allowing for optimization on real hardware.}
\label{fig:parametershift}
\end{figure}

\subsection{Baseline Protocols}
\label{sec:baseline}

We use parallel Ramsey experiments\cite{degen2017quantum} and two realizations of GHZ states \cite{greenberger1989going} as baseline protocols. Ramsey has classical scaling but is more robust, whereas GHZ states achieve high theoretical QFI (Heisenberg scaling) but 1) suffer entanglement gate error and 2) decohere easily. 

Two types of decoder could be constructed for GHZ states, both optimal in the noiseless case, as shown in Figure~\ref{fig:ghz}. We use a comparison of the two as a motivating example to show how two theoretically identical decoders are affected by practical hardware constraints differently. The first decoder is a uniform Hadamard rotation on all qubits which transforms information in parity (GHZ-H), the other one is symmetric to the state prepration circuit - chained CNOT gates followed by a Hadamard gate (GHZ-INV).

\begin{figure}[h]
\centering

GHZ: Uniform-H Decoder
\Qcircuit @C=1em @R=.7em {
     & \gate{H} & \ctrl{1} & \qw    & \qw  & \gate{R_z(\phi)} & \qw & \gate{H} & \qw\\
     & \qw      & \targ    &  \ctrl{1} & \qw & \gate{R_z(\phi)} & \qw & \gate{H} & \qw\\
     & \qw      & \qw      & \targ &  \qw & \gate{R_z(\phi)} & \qw &\gate{H} & \qw
     \gategroup{1}{8}{3}{8}{.7em}{--}
}
\caption*{}
GHZ: Inverse-symmetric Decoder
\Qcircuit @C=1em @R=.7em {
     & \gate{H} & \ctrl{1} & \qw    & \qw  & \gate{R_z(\phi)} & \qw & \qw &\ctrl{1} &\gate{H} & \qw\\
     & \qw      & \targ    &  \ctrl{1} & \qw & \gate{R_z(\phi)} & \qw & \ctrl{1} & \targ &\qw&\qw\\
     & \qw      & \qw      & \targ &  \qw & \gate{R_z(\phi)} & \qw &\targ & \qw & \qw&\qw
     \gategroup{1}{8}{3}{10}{.7em}{--}
}
\caption{Two GHZ decoders. Both are optimal in the noiseless case, but in practical scenarios, they are subject to different noise tradeoffs. The uniform-H decoder avoids higher-error CNOT gates but requires parity readout, which is susceptible to readout errors. The inverse-symmetric decoder uses more CNOT gates, but concentrates information on one qubit, and is thus more robust to readout errors. A comparison of the two shows that different noise combinations favor different decoder designs, and it is not sufficient to only consider QFI of the encoded state.
}
\label{fig:ghz}
\end{figure}

The drawbacks of existing protocols motivate the need to optimize circuits that give relatively high QFI while being robust to system noise/errors, which we demonstrate with variational optimization with a practical objective function.

\section{Variational Optimization}
\label{sec:optimization}
In Section \ref{sec:baseline} we introduced three baseline protocols: parallel Ramsey (independent Ramsey experiments on each sensing qubit), GHZ with uniform-H decoder (GHZ-H), and GHZ with inverse-symmetric decoder (GHZ-INV). These protocols are suboptimal on practical hardware: parallel Ramsey only gives classical (square-root) scaling, and GHZ states tend to decohere easily in noise. Considering imperfect control, interrogation noise, and readout error of each specific platform, we believe an architecture-aware and noise-aware circuit optimization approach can achieve the highest sensitivity under practical constraints. In all our experiments and simulations, we compare our optimization algorithm with the above three baselines.

\subsection{Objective Function}

As mentioned in Section~\ref{sec:CFI}, we base the objective function on CFI. The signal is a $R_z$ rotation on the sensing qubit(s) proportional to time, with angle $\phi=\omega t$. Since we ultimately care about signal $\omega$ rather than $\phi$, this means an extra factor of $t^2$: $\text{CFI}(\omega) = t^2\text{CFI}(\phi)$. Based on Equation~\ref{eqn:cfi_bound}, maximizing SNR per unit time is equivalent to maximizing
\begin{equation}
\sigma_{\hat{\omega}}^{-2} \leq N_{\text{exp}}\cdot \tilde{\text{CFI}}(\omega) = \frac{t^2\text{CFI}(\phi)}{t+t_{\text{overhead}}} \cdot T_{\text{unit}}
\label{eqn:objective}
\end{equation}
where t is signal accumulation time, a parameter we can control, and $t_{\text{overhead}}$ is the time overhead for encoding, decoding and measurement, a parameter determined by hardware and dependent on circuit structure. $N_{\text{exp}}$ is the number of repeated experiments, and $T_{\text{unit}}$ is unit time. We use Equation~\ref{eqn:objective} as the objective function for optimization.

Due to noise, there is a tradeoff between $\text{CFI}(\phi)$ and $t$. Both terms contribute positively to the objective function, yet a large $t$ means longer exposure to interrogation noise, which causes a decrease in $\text{CFI}(\phi)$ ($\text{CFI}(\phi)$ is upper-bounded by $N^2$ for a N-qubit circuit, according to the Heisenberg Limit). We aim to balance this tradeoff by finding a state whose $\text{CFI}(\phi)$ decreases slowly with $t$, meaning the state decoheres slowly in the given noise channel.

The objective function Equation~\ref{eqn:objective} provides a direct bound on SNR and requires a joint optimization of circuit and probe time $t$ in the noisy case.

\subsection{Constraining the Design Space}

The Hilbert space of entangled states is exponentially large. To explore such a space, the first step is to constrain the design space by proposing an ansatz. An ansatz specifies the encoder and decoder structures. Seen from an optimization perspective, an ansatz creates a one-to-one mapping from a parameter vector \bm{$\theta$} to a sensing circuit. Thus, for a fixed ansatz, any circuit metric $C_{\text{\bm{$\theta$}}}$ is simply a function of \bm{$\theta$}, and circuit design becomes a multivariate optimization problem.

The ansatz generation rules are based on both physics principles and hardware capacities. From a physics perspective, since both the initial state and the final measurement are in Z-basis, we stipulate the encoder and decoder structures (though not parameters) to be symmetric. From a co-design perspective, we base our ansatz on the native gates and connectivity graph of hardware. We define three hyperparameters: $l$ circuit layers, $k$ single-qubit gates per layer, and $m$ entanglement gates per layer. By setting these hyperparameters, we could control how local/global we want the entanglement to be, and sequentially explore circuits with increasing depths. Any ansatz that contains redundant, canceling, or spurious gates are discarded. These design choices, while preserving structural flexibility, constrains possible circuit structures to a tractable amount.  

\subsection{Optimization Algorithm}

Combining the components in the sections above, we have the full optimization procedure in Algorithm~\ref{code:opt}. In each iteration, an ansatz is proposed and then optimized with evaluation on real hardware using the ``parameter-shift'' rule.

Given N qubits, available gateset $U$, connectivity graph $G$, and hyper-parameters l (number of layers), k (number of single qubit gates per layer), m (number of two-qubit entanglement gates per layer), the optimization algorithm could be divided into two steps per iteration. The first step is ansatz structure proposal - the algorithm generates random ansatz structures which satisfy platform constraints and create meaningful entanglements. The specific rules for ansatz generation is platform-dependent, and we show an example based on the IBM gateset with our pseudo-code. The proposed ansatz structure parametrizes the circuit into a number of rotation angles, and turns the problem into a continuous multivariate optimization. The second step is continuous optimization. Evaluation of objective function could be done on real hardware using the "parameter-shift" rule, and we choose some classical optimizer (e.g. Powell, COBYLA), considering both the speed of convergence (since hardware evaluation could be expensive) and quality of solution (to avoid local optima).

\begin{algorithm}[h]
\caption{Circuit Learning (with IBM gateset)}
\begin{itemize}[leftmargin=*]
\item[]For $N$-qubit system, given hyper-parameters $k, l, m$
\end{itemize}
\begin{algorithmic}
\FOR{i = 1, ..., $iter_{max}$}

\STATE // Ansatz Construction

\FOR{j = 1, ...$l$}
\STATE 1. Randomly choose $k$ out of $N$ qubits, $q_1, ...q_k$ to apply a single-qubit $U_3$ gate on each. $U_3$ gate on $q_p$ parameterized by ($\theta_{j,p,0}$,$\theta_{j,p,1}$,$\theta_{j,p,2}$) (p=1,...k).

\STATE 2. Based on connectivity graph $G$, randomly choose m non-repeating (control, target) pairs out of $N$ qubits to do an entanglement operation on each pair. If $j=1$, make sure control qubit either 1) has been chosen in step 1, or 2) has been chosen as the target in a prior entanglement operation in the current step.\\
\ENDFOR\\

\STATE // Continuous Optimization
\STATE 1. Initialize $\bm{\theta} \in [0,2\pi]^{6kl}$, $t \leftarrow$ 10$\mu$s (or some value between $\frac{1}{10}T_2$ and $T_2$).

\STATE 2. Use a classical optimizer (e.g. Powell, COBYLA) to obtain  $\max_{{\bm{\theta}},t} C_{\text{\bm{$\theta$}},t} = \frac{t^2\text{CFI}(\phi)}{t+t_{\text{overhead}}}\cdot T_{\text{unit}}$. $\text{CFI}(\phi)$ obtained via $2N+1$ circuit runs using the ``parameter-shift'' rule.

\STATE 3. If $C_{\text{\bm{$\theta$}},t}$ higher than previous max, record optimal {\bm{$\theta$}},t, $C_{\text{\bm{$\theta$}},t}$

\ENDFOR\\

\RETURN optimal {\bm{$\theta$}},t, $C_{\text{\bm{$\theta$}},t}$

\end{algorithmic}
\label{code:opt}
\end{algorithm}

Here we describe the algorithm in more detail using the example of the IBM machine gateset and connectivity. Note that this could be easily generalized to other platforms, detailed in Section~\ref{sec:applications}. The native gateset of IBM machines is comprised of general $U_3$ gates on single qubits (with three free parameters), and CNOT gates for entanglement operations. A combination of $U_3$ gates and CNOT gates per layer allows for the flexibility of creating parallel (or weakly entangled) local structures---for $N$ qubits, if we specify $l=1$ layer, $k=N$ single-qubit gates per layer, and $m=0$ entanglement gates per layer, then a scheme similar to Parallel Ramsey could be recovered. Similarly, if we specify $l=1$ layer, $k=1$ single-qubit gates per layer, and $m=n-1$ entanglement gates per layer, with the correct order of entanglement gates the global GHZ protocol could be recovered. Choosing different values of $l,k,m$, we could obtain arbitrary entangled states. However, a circuit that is too deep will likely incur high gate errors. In our simulations as well as experiments, one layer is usually sufficient.

This algorithm is designed to be generalizable to other sensing platforms and applications. For a different platform, e.g. when we have global twisting interactions which allow us to create spin-squeezed states, we would modify the template-generation part slightly, and no longer aim to create parallel local structures. The continuous optimization part extends naturally to determination of optimal twisting/untwisting, and similar methods are seen in \cite{kaubruegger2019variational, schulte2020ramsey}. In general, for sensing platforms with more global and constrained controls, there will be less flexibility in ansatz construction, but the continuous optimization step becomes more important.

\section{Experiment Results}
\label{sec:expresults}
\begin{figure}[h]
\centering
\includegraphics[width=0.8\columnwidth]{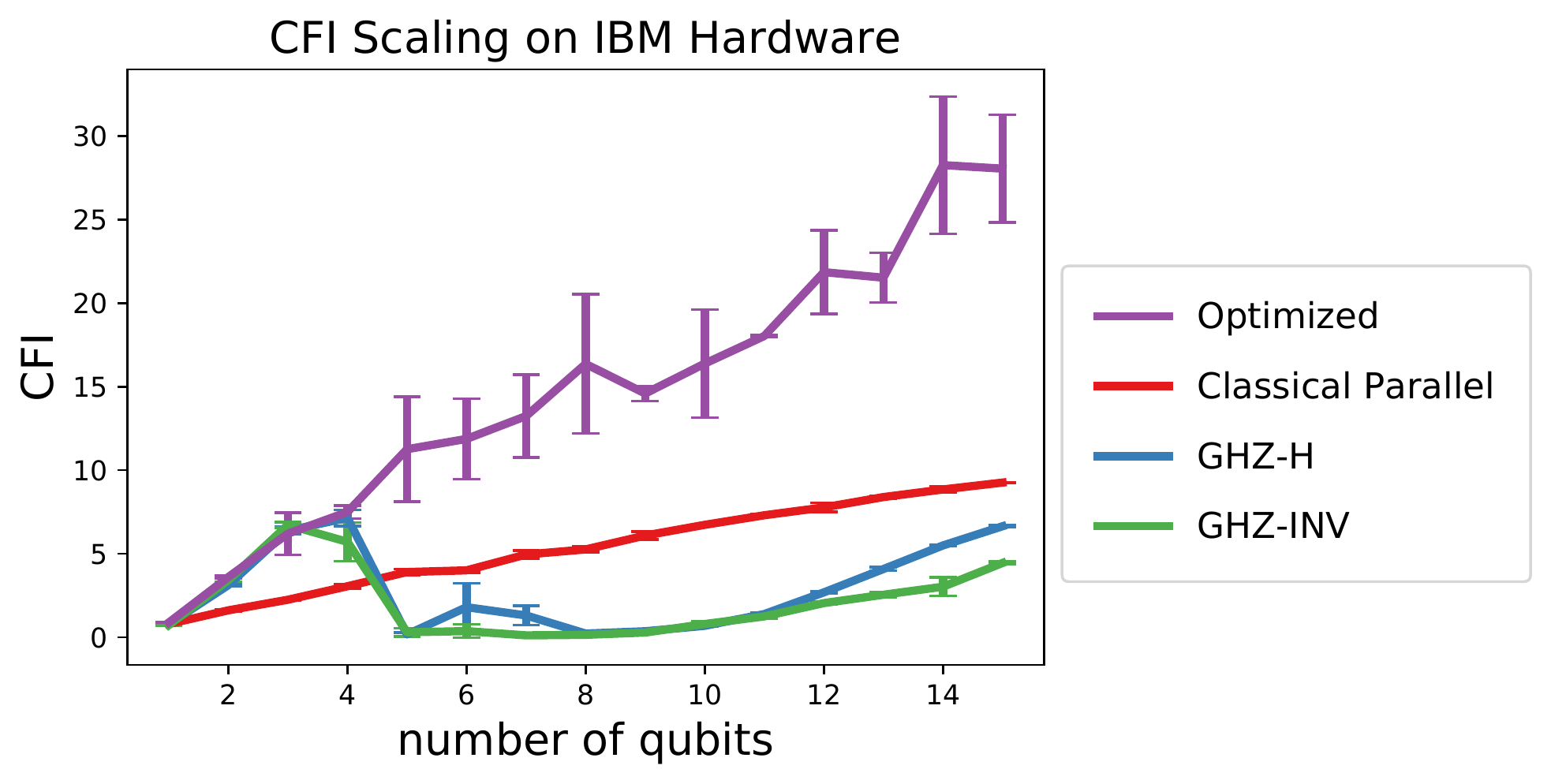}
\caption{CFI scaling of baseline and optimized circuits up to 15 qubits on IBM's machine ``Paris''. Signal is applied with small Rz rotation across the signal accumulation time. The "echo" technique is used (negating the signal and applying X) to suppress detuning, as in other sensing applications. GHZ protocols yield decreasing performance at $\geq$ 4 qubits. The optimized circuit employs local entangled structures and overcomes this problem, achieving up to 3.19x CFI gain from the classical limit, and up to 13.2x SNR improvement from GHZ protocol. Due to limited hardware access, the the number of total optimization iterations and repeats we can run is limited. For this reason, we fix sensing time to be 0.355us in experiment (whereas in simulation we are able to co-optimize time, resulting in longer $t$ and lower $\text{CFI}(\phi)$, seen in Figure~\ref{fig:full_noise}). The variance can also be expected to be smaller without such limitations.}
\label{fig:exp}
\end{figure}

We demonstrate the optimization algorithm on IBM quantum computer ``Paris'', up to 15 qubits. Signal is applied with small Rz rotations uniformly across all sensing qubits throughout the signal accumulation time. We note that for such a system size, simulation-based optimization is infeasible due to the expensive computation of simulation of 15-qubit circuits (which needs to be repeated thousands of times as an optimization subroutine). Optimization outperforms all existing baseline protocols, achieving up to 3.19x gain from the classical limit.
%In experiment, we specify a fixed interrogation time of 0.355$\mu$s and use the echo technique to cancel detuning error (applying X gate and reverting signal at midpoint \cite{maze2008nanoscale}).
Furthermore, we note that the baseline protocols are run with existing circuit compiler optimization to unnecessary swap gates based on hardware connectivity map, which shows the advantage of variational circuit optimization compared to existing (fixed) compiler optimization.

%Each optimization run takes approximately 1 hour of runtime on IBM machine (for N qubits, $8192\times(2N+1)$ repeated circuit evaluations per optimization step within one run). One round of experiment for 1-5 qubits takes around 5 hours of runtime, which spans between 1 and 5 days depending on queue time. Due to the monthly quota of hardware access, it is unrealistic to perform a hundred independent experiments. We have been able to perform six repetitions, and the error bars in \ref{fig:exp} plot the max and min values of experiment runs. A larger number of repetitions will likely give a tighter confidence interval, but from current results we see that as system size grows, even the min value from optimization runs clearly outperforms existing protocols.

%Due to the high queueing time of experiment runs and time-variability of the machine, optimization done on real hardware is much less efficient than in simulation and hard to converge. Due to these practical constraints, in order to limit on-device optimization to finish within a reasonable time, we first run optimizations in simulation and initialize on-device optimization with simulation results. The COBYLA optimizer \cite{powell2007view} is chosen for its efficiency in objective function evaluations. Considering the difficulty of convergence due to queue time and machine variability, the maximum values across optimization iterations are chosen as the final output. 

We observe that for GHZ states, CFI starts to decrease when we have more than three qubits, probably due to errors from the chained CNOT gates and the short coherence time of multi-qubit GHZ states. This might explain the significant drop in performance of GHZ states (vastly underperforming classical parallel) for more than four qubits.

While analyzing the result states returned by the adaptive learning protocol, we noticed that variational optimization never returned max entanglement chain of length greater than 5, and the circuit structure is generally not symmetric across probe qubits---by taking advantage of local entanglements and adapting to variation in qubit/gate qualities, adaptive optimization achieves higher scalability. The circuit learning approach enables more flexibility in balancing the tradeoffs between a large number of entangled qubits (higher QFI capacity) vs. gate noise from entanglement operations, as well as the sensitivity of strongly-entangled states such as GHZ vs. their susceptibility to noise. The effects of different noise sources are studied in more detail in Section \ref{sec:simresults}.

We also noticed the difference between experiment and simulation with device statistics at long accumulation time (Figure \ref{fig:exp} and Figure \ref{fig:full_noise})--- although simulation suggests signal accumulation time could be much longer (to maximize SNR), in practice we observed that results can be quite inconsistent across repeated trials when accumulation time is long, probably due to other fluctuating noise sources on the real device. Due to this reason, accumulation time in experiment is kept much shorter than simulation, resulting in more significant effects of gate/read noise as opposed to T1/T2 noise in experiment.

\section{Optimization Results under Different Platform Conditions}
\label{sec:simresults}
\subsection{Optimization with Different Noises}

We demonstrate that our optimization algorithm is system-adaptive by simulating different platform conditions, including different noise decompositions, signal distributions, and gateset or connectivity constraints.

\subsubsection{Noise Decomposition}

Noise is incurred by each operation---encoding, signal accumulation, decoding, and readout. The contribution of each noise component depends on specific circuit parameters (encoder/decoder depth, accumulation time, whether the output distribution is ``concentrated'' on some qubits, etc.).
As mentioned in Section~\ref{sec:CFI}, QFI quantifies the amount of information contained in the quantum state and decreases monotonically under noisy quantum channels. Calculating the QFI at different points in the circuit allows us to see how much information is lost at each (noisy) step, as shown in Figure~\ref{fig:decompose_noise}. We compare 4 circuits: a circuit obtained via optimization with full noise and three baseline baselines- Parallel Ramsey, GHZ with uniform-H decoder, and GHZ with inverse-symmetric decoder. We observe that parallel Ramsey starts with low QFI but is more robust to noise, whereas GHZ states start with high QFI (optimal in noiseless case), but decohere easily during interrogation. The optimized result starts with a QFI lower than GHZ, but is more robust to noise, yielding a high QFI after all noisy operations. 

\begin{figure*}[h]
\renewcommand{\arraystretch}{1.37}
\centering
\begin{minipage}{0.7\columnwidth}
    \includegraphics[width=1\columnwidth]{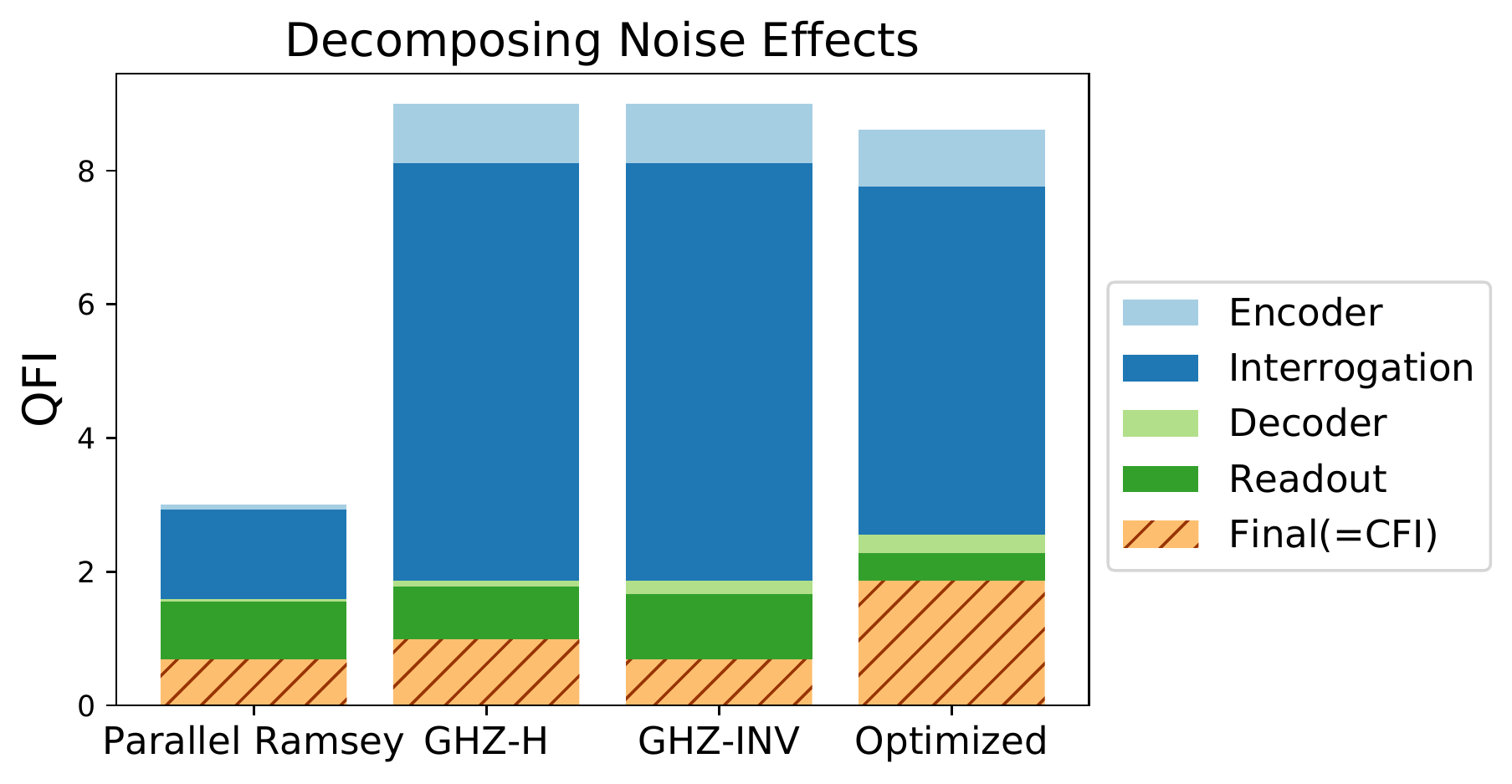}
  \end{minipage}
  \begin{minipage}{1\columnwidth}\small
      \centering
      \begin{tabular}
      {p{0.18\columnwidth}|p{0.23\columnwidth}|p{0.45\columnwidth}}
        \multicolumn{3}{c}{\textbf{Baseline Protocol Tradeoffs}} \\
        Protocol & Advantage & Disadvantage \\
        \hline \hline
        Parallel Ramsey & Robust & Only square root scaling \\
        \hline
        GHZ-H &Linear scaling if no noise & Decoheres quickly, suceptible to readout noise \\
        \hline
        GHZ-INV &  Linear scaling if no noise & Decoheres quickly, susceptible to gate noise\\
        \hline
      \end{tabular}
\end{minipage}

\caption{Decomposition of noise effects from each step: encoder, interrogation, decoder, readout. Although these noise sources are independent, their effects need to be considered holistically---a state that is robust to one noise type usually is vulnerable to others. Optimization aims to balance the tradeoff of these noise effects. The orange (bottom) part shows the final QFI after all noises, equal to CFI, which is our optimization objective. Each colored region shows the additional QFI improvement of given circuit \textbf{if} the labeled operation is perfect. Total height denotes QFI in the ideal case. The optimized circuit has less total QFI than GHZ states because we are optimizing for the highest practical (orange) region. Our optimization found a protocol that performs 2-3x better specifically under the given constraints. This plot is based on a simulation of a 3-qubit sensor with a 10kHz signal and a fixed probing time of 20$\mu$s.
}
\label{fig:decompose_noise}
\end{figure*}

\subsubsection{Optimization Results} 

\begin{figure*}[h]
\centering
\includegraphics[width=1.58\columnwidth]{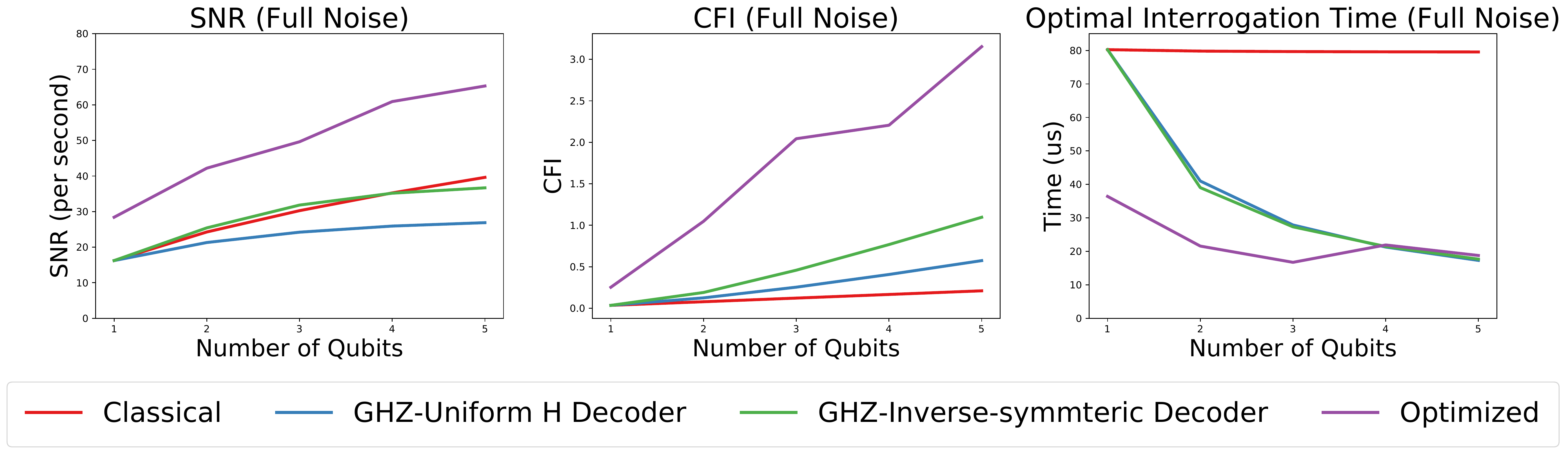}
\caption[Caption for LOF]{Optimization results with full noise. Noise statistics are based average of hardware calibration data of IBM superconducting hardware: single qubit 1\% depolarizing channel, CNOT 3\% (independent) depolarizing channels on both qubits, readout 5\% error, $T_1$ 52.2$\mu$s, $T_2$ 62.8$\mu$s. Signal strength is 10kHz. The optimization objective is $\frac{t^2\text{CFI}(\phi)}{t+t_{\text{overhead}}}$ (Equation~\ref{eqn:objective}) with a hardware-determined $t_{\text{overhead}}$. Optimization requires balancing the tradeoff between CFI and interrogation time $t$. In addition to SNR derived from our objective function, we also plot out the optimal CFI and $t$ which maximize the optimization objective.
}
\label{fig:full_noise}
\end{figure*}
In this section, we consider four different noise scenarios: full noise model including gate noise, interrogation noise, and readout noise, and removing (or suppressing) each noise source. In each noisy scenario, we run our optimization algorithm to maximize SNR and compare the optimized results with baseline sensing protocols: parallel Ramsey, GHZ state with uniform-H decoder, GHZ state with inverse-symmetric decoder. The optimization target is Equation~\ref{eqn:objective}. For each baseline circuit, we only optimize on the interrogation time parameter. For the optimized result, we jointly optimize circuit and interrogation time.

For optimization, we use Powell \cite{powell1964efficient} and COBYLA \cite{powell2007view} gradient-free optimizers. Powell allows for the exploration of a relatively large parameter space and is less likely to return a local maximum in our optimization application, although at the cost of a larger number of iterations. COBYLA is observed to converge much faster, although sometimes outputs local maxima. The simulation results shown in this section come from Powell optimizer which in general finds better solutions, but in experiment (or for large simulation) where circuit evaluations are more expensive, we use COBYLA for its evaluation efficiency.

The orange (bottom) part shows the final QFI after all noises, equal to CFI, which is our optimization objective. Each colored region shows the additional QFI improvement of given circuit \textbf{if} the labeled operation is perfect. Total height denotes QFI in the ideal case. The optimized circuit has less total QFI than GHZ states because we are optimizing for the highest practical (orange) region. A comparison between GHZ-H and GHZ-INV shows that decoder noise is higher in GHZ-INV, which is expected since GHZ-INV has 2 CNOT gates in decoder, whereas GHZ-H only has single-qubit gates. The readout noise in GHZ-INV is expected to be smaller than GHZ-H when the starting state is somewhat close to GHZ. We suspect the stronger effect from readout noise to GHZ-INV here is because the final state is already highly decohered, thus making the effect of readout noise harder to predict. Overall, our adaptive optimization found a protocol that performs 2-3x better specifically under the given constraints. 

As shown in Figure~\ref{fig:full_noise}, under full noise, the classical parallel circuit allows for the longest probe time, although with a relatively low CFI. GHZ states give higher CFI but are limited to a much shorter probe time. The parallel scheme gives square root scaling, and the scaling of GHZ states approximately saturates. It is also notable that GHZ with an inverse-symmetric decoder achieves much higher sensitivity than the uniform-H decoder, which shows that readout errors incurred by the parity-readout for the uniform-H decoder outweigh the errors from the $N-1$ extra CNOT gates of the inverse-symmetric decoder.

The optimized circuit achieves much better scaling than baseline circuits. Theoretically, when we have depolarizing channels which overlap with the signal direction, the asymptotic scaling cannot surpass square root \cite{zhou2018achieving}. Although we are still in the few-qubit regime and far from asymptotics, this might explain why we could not get a linear scaling.

Optimization yields up to 1.75x SNR gain in 5 qubits. The optimized circuit obtains higher SNR largely by achieving a high CFI, at the cost of a relatively short probe time. The optimized results all come from one-layer ansatz's. For 1 to 3 qubits, the optimized results entangle the full system, whereas for 4-qubit and 5-qubit systems the optimal solutions are two entangled subsystems. The reason is probably twofold: 1) gate errors during encoding/decoding accumulate when we entangle too many qubits; 2) Strongly entangled states like GHZ states are highly susceptible to noise and is no longer optimal for sensing under a non-negligible amount of noise. The high noise susceptibility leaves us with a shorter probe time which hurts sensitivity. GHZ state can be seen as an extreme case of both these effects. The optimized circuit, by leveraging local rather than global entanglements, is able to overcome these issues and achieve better scalability.

\begin{figure*}[h]
\centering
\includegraphics[width=1.58\columnwidth]{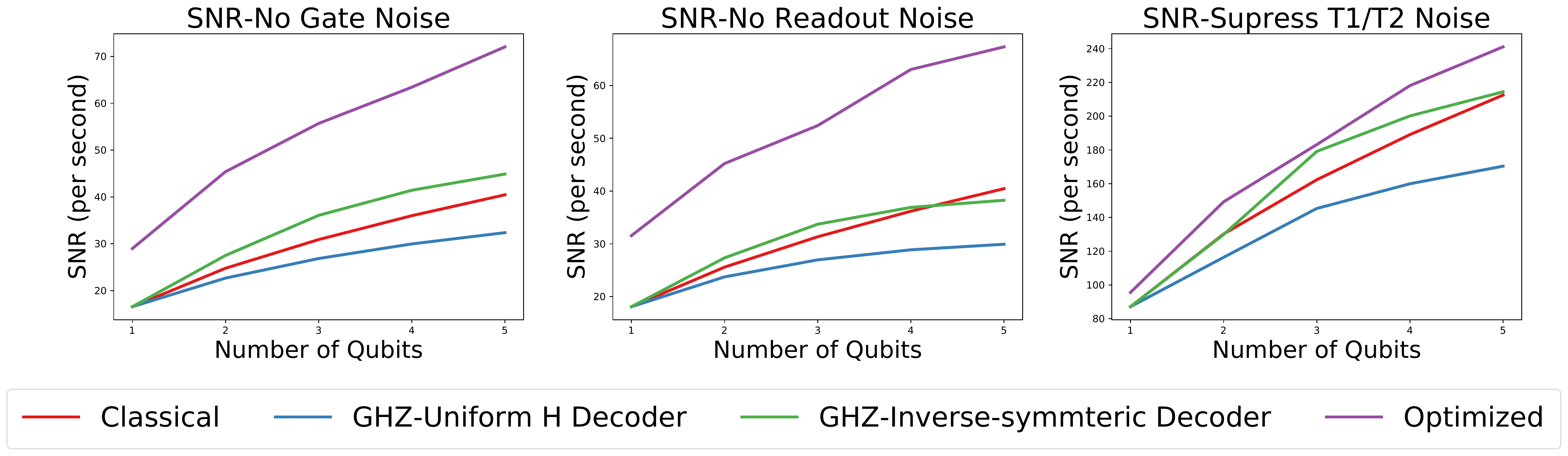}
\caption{Optimization results removing/suppressing each noise source. Optimized circuit outperforms all existing protocols in each scenario, yielding up to 1.75x, 1.74x, 1.15x SNR improvement respectively. Note that the advantage of optimization more pronounced when we have $T_1$/$T_2$ noise (compared to suppressing them), which shows that out optimization method is effective in finding circuits that are more robust to T1/T2 noise.
}
\label{fig:remove_each}
\end{figure*}

To better understand the effects of each noise source, we also removed/suppressed each noise type, keeping the other two sources constant. Gate and readout noises can be removed by simulating perfect gates/readout. Interrogation noise is \textit{decoherence} noise coming from the characteristic lifetimes of the qubit, called $T_1$ and $T_2$. Shorter interrogation times (relative to $T_1$ and $T_2$) lead to lower \textit{decoherence} error. To suppress interrogation noise, (Note that we could not remove interrogation ($T_1$,$T_2$) noise entirely, since that will result in infinite probing time and thus infinite sensitivity.)
we simulate $T_1$ and $T_2$ as 10 times their actual value. Results are shown in Figure~\ref{fig:remove_each}. We obtain 1.75x, 1.74x, 1.15x SNR improvements when removing gate noise, removing readout noise, and suppressing $T_1$/$T_2$ noise, respectively.

Note that GHZ with inverse-symmetric decoder outperforms the classical scheme once we remove gate noise, which shows the effects of CNOT gate errors.
What is surprising at first sight is that after removing readout error, inverse-symmetric decoder still performs better than uniform-H decoder for GHZ states, even with the $N-1$ extra CNOT gates. A closer QFI analysis shows that although an H rotation before measurement is optimal in the noiseless case, with interrogation noise H is no longer the best rotation direction. Rotating each qubit in a suboptimal direction, in this case, results in less system QFI compared to a chain of CNOT gates followed by a rotation of one qubit in a suboptimal direction.
The advantage of optimization becomes smaller if we suppress interrogation errors while keeping gate and readout error constant. This shows our optimization mainly targets interrogation noise, which is the largest noise component shown in Figure~\ref{fig:decompose_noise}. When interrogation noise is suppressed, the dominant noise source comes from the control itself, and simple protocols are highly favored. Our optimization technique provides the greatest gain when the control noise is not highly dominant, meaning we can afford to create interesting (and probably slightly more complex) states.

\subsection{Optimization with different signal distributions}
\begin{figure}[h]
\centering
\includegraphics[width=0.8\columnwidth]{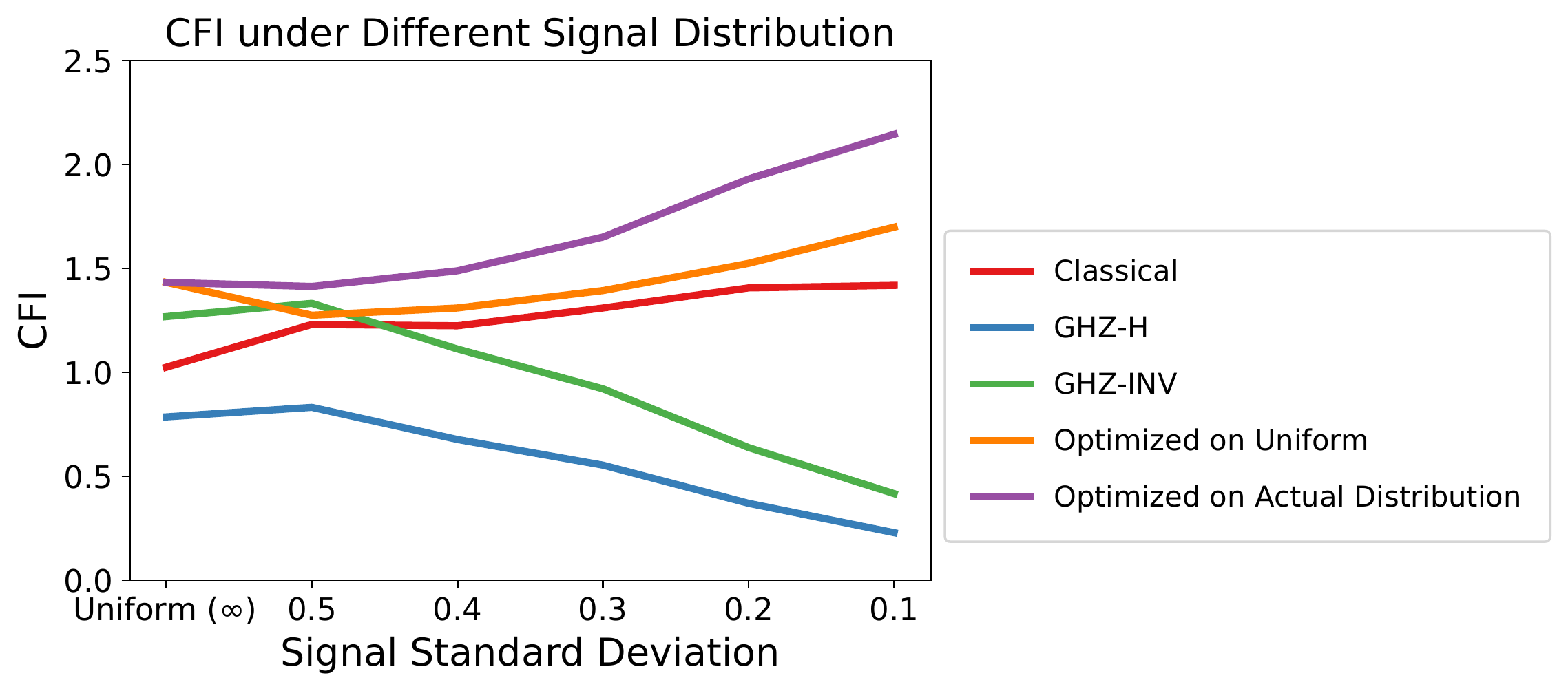}
\caption{Optimization results under different signal distributions. Since the signal is non-constant, we use CFI rather than the signal-dependent SNR as the metric (see Section~\ref{sec:CFI}). The data is taken from a simulation of three qubits under the full noise model. Signal (Rz rotation angles) are modeled as Gaussians with unit mean and varying standard deviation, as displayed on the x-axis. Note that although we vary the distribution of signal magnitude, the signal is always applied uniformly to each sensing qubit (i.e. the Rz rotation angles on qubits i and j are kept the same). We compare baseline protocols, a circuit optimized on uniformly-distributed signal (``Optimized on Uniform'', orange), and a circuit optimized on the actual signal distribution (``Optimized on Actual'', purple). Optimizing on actual data gives the highest sensitivity, which shows that our optimization technique can take advantage of non-uniform signal distributions, yielding up to 1.51x CFI improvement. We also observe that optimizing with different distributions yields the same circuit structure, but different circuit parameters.
}
\label{fig:signal_distrib}
\end{figure}

Different sensing applications concern input signals with different distributions. By adaptively optimizing on the actual signal, our optimization algorithm can exploit the non-uniform signal distribution and find a circuit that works especially well with the given distribution. In this section, we compare baseline circuits with optimization run on uniform as well as Gaussian signal distributions with different variances. We observe that optimizing with different distributions yields the same circuit structure, but different circuit parameters. As shown in Figure~\ref{fig:signal_distrib} optimization run on non-uniform signal distributions gives the best results (up to 1.51x CFI improvement), with increasing gain at small variances. This is especially well-suited to applications where the signal has frequency peaks, such as NMR spectroscopy \cite{abraham1998introduction}.

\subsection{Optimization with limited control}
The other important flexibility given by our optimization technique is that we could tailor to the available control of each platform. On some platforms, the control is greatly limited. Since our optimization technique takes gateset and connectivity map as inputs and proposes ansatz's based on these constraints, it is also suitable for such applications.

\subsubsection{Collective Control}
For certain sensing applications, e.g. dark spins of nitrogen-vacancy (NV) platform \cite{goldstein2011environment} (detailed in Section~\ref{sec:nv}), we do not have individual addressability of the probe spins. Each reporter spin will undergo roughly the same transformations given the drive field, and thus the available control is an arbitrary rotation uniformly applied to all probes. Likewise, the $S_z$-$S_z$ entanglement operation between NV and probes also applies simultaneously to all probes, which could be modeled as simultaneous controlled-$R_z$ gates with different rotation angles, and thus easily interface into our optimization framework.

\subsubsection{Limited Connectivity}
On certain platforms, the qubit connectivity is severely limited. Although in principle, by ``relaying'' operations we are able to perform entanglement on an arbitrary pair of qubits, the gate error and time overhead could be considerable, and thus a good model should take the connectivity constraints into consideration. For example, nuclear spins around the NV center \cite{bradley2019ten} could only entangle with each other via the NV center. Such connectivity constraints are entirely handled by our optimization framework in Algorithm~\ref{code:opt} which specifies the ansatz based on the connectivity map.

\section{Outlook}
\label{sec:outlook}
\subsection{Applications}
\label{sec:applications}
Based on the variational optimization method's capability of achieving high sensing performance while considering noise, gate error, and limited control, our method could be widely applied to different sensing platforms and applications.

\subsubsection{Nitrogen-Vacancy Sensing}
\label{sec:nv}
The nitrogen-vacancy (NV) center in diamond is a promising sensing platform due to its
long coherence time in room temperature \cite{bar2013solid} and photoluminescence property, which allows for easy initialization and readout \cite{doherty2013nitrogen}. NV centers can be used to sense magnetic field \cite{boss2017quantum,barry2020sensitivity}, electric field \cite{dolde2011electric} and temperature \cite{neumann2013high,kucsko2013nanometre,clevenson2015broadband}. 
Surface reporter spins \cite{sushkov2014magnetic,schaffry2011proposed} and NV ensemble sensing \cite{glenn2018high} show promise in increasing sensitivity and reducing sensing volume. Circuit learning could be used to exploit probe spins near the diamond surface for sensitivity enhancement. Abstracted to a sensing circuit, this corresponds to a constrained gateset of a time-dependent probes-NV entanglement gate and uniform controls on probe spins. 
Via adaptive optimization, we could adjust the circuit to the material-determined distribution of probe spins, utilize the dipolar-dipolar interaction between the probe spins, and optimize the orientation of the sensing state as well as entanglement duration.

\subsubsection{Dark Matter Detection}
%Dark Matter is one of the most important unsolved problems in physics;
About 85\% of the matter in today's universe consists of dark matter and dark energy \cite{sahni20045}. 
\textit{Axions} are one of the most promising dark matter candidates \cite{preskill1983cosmology,sikivie1983experimental}, and can be detected by the resonance frequency shift of the qubit coupling to the superconducting cavity \cite{dixit2018detecting}. However, axion detection is difficult since long-distance entanglement is required \cite{pospelov2013detecting}---one main challenge is photon loss in long-distance fiber \cite{safronova2018search}. Our method could help alleviate photon loss by optimizing for the optimal entangled photon state in spatially separated superconducting cavity detectors \cite{ge2018distributed}.

\subsubsection{Squeezed States}
Squeezed states improve sensitivity by suppressing the uncertainty on one quadrature observable while amplifying the uncertainty over another non-commutative quadrature observable. Experimental progress includes LIGO for gravitational wave detection \cite{aasi2013enhanced,yu2020quantum} and spin-squeezing using cold atoms for spectroscopy \cite{schleier2010states,appel2009mesoscopic,braverman2019near}. Building upon previous works \cite{kaubruegger2019variational,schulte2020ramsey}, our method can be used to optimize for twisting time, signal direction, and measurement scheme for higher sensitivity.

\subsubsection{Atomic Clocks}

Atomic clocks are currently used to define a second with the precision of $10^{-18}$ (which means it will err by 1 second within the age of our universe) \cite{bothwell2019jila}, and play a central role in precision measurement applications such as global positioning system (GPS). Finding and using various techniques to keep the atomic transition stable is crucial to sensitivity \cite{ludlow2015optical}. Atomic clocks use a feedback loop to lock the frequency of a local oscillating field to the atomic transition frequency, and the oscillating field frequency is defined as the reference clock. Our circuit optimization can be applied to the feedback loop system to potentially stabilize the atomic transition and the laser frequency beyond the current limitations.

\subsection{Future Directions}
We demonstrate an optimization algorithm which surpasses the classical limit and overcomes the plateauing or decreasing performance of existing entanglement-based protocols at increased system sizes. We believe the algorithm could be further improved by considering the following areas.

One of the challenges of running optimization on hardware is the difficulty of obtaining convergence since device noise fluctuates with time, and the optimal circuit at each given time is different. For this reason, it is key that the optimization routine does not take too long (so that the noise does not fluctuate too much), and that the optimized sensing circuit should be updated periodically. This means an optimizer that is efficient in the number of objective function evaluations is highly important. In our study, we did a preliminary comparison of different optimizers, including Nelder-Mead \cite{nelder1965simplex}, Powell \cite{powell1964efficient}, COBYLA \cite{powell2007view}, and in general observed a tradeoff between speed of convergence and solution quality. The simulation results in this paper are based on Powell, which converges slowly but outputs high-quality solutions, whereas experiment results are based on COBYLA which is more evaluation-efficient (by doing polynomial interpolation with points explored in the parameter space) but sometimes outputs suboptimal solutions. For the optimization algorithm to be run in real sensing applications, we hope to find a robust optimizer which strikes a good balance of convergence speed and solution quality.

Another way to speed up the optimization is to reduce the number of proposed ansatz's. The advantage of our algorithm comes largely from the flexibility of constructing various circuit structures based on available gateset and connectivity. However, without losing this flexibility, we would benefit from automatically ``ruling out'' suboptimal ansatz's, which would be possible if we design some heuristics to guide ansatz selection. The current algorithm does rule out some obviously bad circuit structures, but by considering permutation symmetry, circuit transformation/simplification, we could make the template construction step smarter, thereby saving optimization cost.

We also hope to investigate whether having ancilla qubits that are not exposed to the signal could help improve sensitivity. This is relevant in some sensing applications, such as the NV sensing platform, where we have control over nuclear spins close to the NV center \cite{bradley2019ten}.

Finally, to make this optimization more applicable to real sensing applications, we would like to consider more realistic physical models, such as interactions between probe qubits. We would also like to consider more complex and practical signal distributions, for example in NMR spectroscopy applications. Our adaptive circuit optimization could also fit into a Bayesian framework to fully take advantage of prior knowledge of the signal.

\section{Conclusions}
Quantum sensing is an emerging area of quantum technology, which exploits quantum mechanical effects on different hardware platforms to achieve enhanced sensitivity. Quantum sensing already has practical impacts in areas such as magnetometry \cite{fagaly2006superconducting} and timekeeping \cite{ludlow2015optical}. Recent experimental progress such as LIGO demonstrates the power of entanglement-enhanced sensing \cite{aasi2013enhanced,yu2020quantum}. We are entering the exciting transition from proof-of-concept experiments to practical applications \cite{pezze2018quantum}.

On practical hardware, where we have limited control capacity over a relatively small number of qubits and noise from various sources, we find existing protocols that are optimal in ideal cases to perform poorly. Thus, it is important to adopt a co-design approach: find sensing circuits tailored to the specific underlying hardware and exploit application-specific noise and signal characteristics. We demonstrate a quantum-classical hybrid algorithm for sensing circuit optimization. Our approach mirrors the myriad of recent work \cite{penney2019survey} on applying machine learning approaches to improve classical system optimization and design optimization. Our optimization algorithm surpasses the classical limit and overcomes the plateauing/decreasing performance of existing protocols, both in simulation and in experiment. We also show that the algorithm is especially advantageous when trained on a highly non-uniform signal distribution, which applies to various sensing applications such as NMR spectroscopy. The optimization method could easily be generalized to different platforms with different types of control, even going beyond the gate model abstraction and directly parameterizing the control Hamiltonians. We propose extensions of our method for different sensing applications range from nanoscale field sensing to dark matter detection and timekeeping. We believe that a systems design approach with noise-awareness, realistic classical-quantum interfacing, and adaptivity to diverse physical platforms and applications is essential to achieving quantum advantage on practical sensing hardware.

%\section*{Acknowledgment}

%The preferred spelling of the word ``acknowledgment'' in America is without an ``e'' after the ``g''. Avoid the stilted expression ``one of us (R. B. G.) thanks $\ldots$''. Instead, try ``R. B. G. thanks$\ldots$''. Put sponsor acknowledgments in the unnumbered footnote on the first page.
\bibliographystyle{unsrt}
\bibliography{refs}

\begin{thebibliography}{10}

\bibitem{pezze2018quantum}
Luca Pezze, Augusto Smerzi, Markus~K Oberthaler, Roman Schmied, and Philipp
  Treutlein.
\newblock Quantum metrology with nonclassical states of atomic ensembles.
\newblock {\em Reviews of Modern Physics}, 90(3):035005, 2018.

\bibitem{ludlow2015optical}
Andrew~D Ludlow, Martin~M Boyd, Jun Ye, Ekkehard Peik, and Piet~O Schmidt.
\newblock Optical atomic clocks.
\newblock {\em Reviews of Modern Physics}, 87(2):637, 2015.

\bibitem{leibfried2004toward}
D~Leibfried, Murray~D Barrett, T~Schaetz, J~Britton, J~Chiaverini, Wayne~M
  Itano, John~D Jost, Christopher Langer, and David~J Wineland.
\newblock Toward heisenberg-limited spectroscopy with multiparticle entangled
  states.
\newblock {\em Science}, 304(5676):1476--1478, 2004.

\bibitem{dixit2018detecting}
Akash Dixit, Aaron Chou, and David Schuster.
\newblock Detecting axion dark matter with superconducting qubits.
\newblock In {\em Microwave Cavities and Detectors for Axion Research}, pages
  97--103. Springer, 2018.

\bibitem{abobeih2019atomic}
MH~Abobeih, J~Randall, CE~Bradley, HP~Bartling, MA~Bakker, MJ~Degen, M~Markham,
  DJ~Twitchen, and TH~Taminiau.
\newblock Atomic-scale imaging of a 27-nuclear-spin cluster using a quantum
  sensor.
\newblock {\em Nature}, 576(7787):411--415, 2019.

\bibitem{farhi2014quantum}
Edward Farhi, Jeffrey Goldstone, and Sam Gutmann.
\newblock A quantum approximate optimization algorithm.
\newblock {\em arXiv preprint arXiv:1411.4028}, 2014.

\bibitem{degen2017quantum}
Christian~L Degen, F~Reinhard, and Paola Cappellaro.
\newblock Quantum sensing.
\newblock {\em Reviews of modern physics}, 89(3):035002, 2017.

\bibitem{giovannetti2004quantum}
Vittorio Giovannetti, Seth Lloyd, and Lorenzo Maccone.
\newblock Quantum-enhanced measurements: beating the standard quantum limit.
\newblock {\em Science}, 306(5700):1330--1336, 2004.

\bibitem{giovannetti2011advances}
Vittorio Giovannetti, Seth Lloyd, and Lorenzo Maccone.
\newblock Advances in quantum metrology.
\newblock {\em Nature photonics}, 5(4):222, 2011.

\bibitem{toth2014quantum}
G{\'e}za T{\'o}th and Iagoba Apellaniz.
\newblock Quantum metrology from a quantum information science perspective.
\newblock {\em Journal of Physics A: Mathematical and Theoretical},
  47(42):424006, 2014.

\bibitem{schleier2010states}
Monika~H Schleier-Smith, Ian~D Leroux, and Vladan Vuleti{\'c}.
\newblock States of an ensemble of two-level atoms with reduced quantum
  uncertainty.
\newblock {\em Physical review letters}, 104(7):073604, 2010.

\bibitem{appel2009mesoscopic}
J{\"u}rgen Appel, Patrick~Joachim Windpassinger, Daniel Oblak, U~Busk Hoff,
  Niels Kj{\ae}rgaard, and Eugene~Simon Polzik.
\newblock Mesoscopic atomic entanglement for precision measurements beyond the
  standard quantum limit.
\newblock {\em Proceedings of the National Academy of Sciences},
  106(27):10960--10965, 2009.

\bibitem{aasi2013enhanced}
The LIGO~Scientific Collaboration.
\newblock Enhanced sensitivity of the ligo gravitational wave detector by using
  squeezed states of light.
\newblock {\em Nature Photonics}, 7(8):613--619, 2013.

\bibitem{yu2020quantum}
Haocun Yu, L.~McCuller, M.~Tse, L.~Barsotti, N.~Mavalvala, J.~Betzwieser, C.~D.
  Blair, S.~E. Dwyer, A.~Effler, M.~Evans, A.~Fernandez-Galiana, P.~Fritschel,
  V.~V. Frolov, N.~Kijbunchoo, F.~Matichard, D.~E. McClelland, T.~McRae,
  A.~Mullavey, D.~Sigg, B.~J.~J. Slagmolen, C.~Whittle, A.~Buikema, Y.~Chen,
  T.~R. Corbitt, R.~Schnabel, R.~Abbott, C.~Adams, R.~X. Adhikari, A.~Ananyeva,
  S.~Appert, K.~Arai, J.~S. Areeda, Y.~Asali, S.~M. Aston, C.~Austin, A.~M.
  Baer, M.~Ball, S.~W. Ballmer, S.~Banagiri, D.~Barker, J.~Bartlett, B.~K.
  Berger, D.~Bhattacharjee, G.~Billingsley, S.~Biscans, R.~M. Blair, N.~Bode,
  P.~Booker, R.~Bork, A.~Bramley, A.~F. Brooks, D.~D. Brown, C.~Cahillane,
  K.~C. Cannon, X.~Chen, A.~A. Ciobanu, F.~Clara, S.~J. Cooper, K.~R. Corley,
  S.~T. Countryman, P.~B. Covas, D.~C. Coyne, L.~E.~H. Datrier, D.~Davis, C.~Di
  Fronzo, K.~L. Dooley, J.~C. Driggers, P.~Dupej, T.~Etzel, T.~M. Evans,
  J.~Feicht, P.~Fulda, M.~Fyffe, J.~A. Giaime, K.~D. Giardina, P.~Godwin,
  E.~Goetz, S.~Gras, C.~Gray, R.~Gray, A.~C. Green, Anchal Gupta, E.~K.
  Gustafson, R.~Gustafson, J.~Hanks, J.~Hanson, T.~Hardwick, R.~K. Hasskew,
  M.~C. Heintze, A.~F. Helmling-Cornell, N.~A. Holland, J.~D. Jones,
  S.~Kandhasamy, S.~Karki, M.~Kasprzack, K.~Kawabe, P.~J. King, J.~S. Kissel,
  Rahul Kumar, M.~Landry, B.~B. Lane, B.~Lantz, M.~Laxen, Y.~K. Lecoeuche,
  J.~Leviton, J.~Liu, M.~Lormand, A.~P. Lundgren, R.~Macas, M.~MacInnis, D.~M.
  Macleod, G.~L. Mansell, S.~Márka, Z.~Márka, D.~V. Martynov, K.~Mason, T.~J.
  Massinger, R.~McCarthy, S.~McCormick, J.~McIver, G.~Mendell, K.~Merfeld,
  E.~L. Merilh, F.~Meylahn, T.~Mistry, R.~Mittleman, G.~Moreno, C.~M.
  Mow-Lowry, S.~Mozzon, T.~J.~N. Nelson, P.~Nguyen, L.~K. Nuttall, J.~Oberling,
  Richard~J. Oram, C.~Osthelder, D.~J. Ottaway, H.~Overmier, J.~R. Palamos,
  W.~Parker, E.~Payne, A.~Pele, C.~J. Perez, M.~Pirello, H.~Radkins, K.~E.
  Ramirez, J.~W. Richardson, K.~Riles, N.~A. Robertson, J.~G. Rollins, C.~L.
  Romel, J.~H. Romie, M.~P. Ross, K.~Ryan, T.~Sadecki, E.~J. Sanchez, L.~E.
  Sanchez, T.~R. Saravanan, R.~L. Savage, D.~Schaetzl, R.~M.~S. Schofield,
  E.~Schwartz, D.~Sellers, T.~Shaffer, J.~R. Smith, S.~Soni, B.~Sorazu, A.~P.
  Spencer, K.~A. Strain, L.~Sun, M.~J. Szczepańczyk, M.~Thomas, P.~Thomas,
  K.~A. Thorne, K.~Toland, C.~I. Torrie, G.~Traylor, A.~L. Urban, G.~Vajente,
  G.~Valdes, D.~C. Vander-Hyde, P.~J. Veitch, K.~Venkateswara, G.~Venugopalan,
  A.~D. Viets, T.~Vo, C.~Vorvick, M.~Wade, R.~L. Ward, J.~Warner, B.~Weaver,
  R.~Weiss, B.~Willke, C.~C. Wipf, L.~Xiao, H.~Yamamoto, Hang Yu, L.~Zhang,
  M.~E. Zucker, and J.~Zweizig.
\newblock Quantum correlations between the light and kilogram-mass mirrors of
  ligo.
\newblock {\em arXiv preprint arXiv:2002.01519}, 2020.

\bibitem{koczor2019variational}
B{\'a}lint Koczor, Suguru Endo, Tyson Jones, Yuichiro Matsuzaki, and Simon~C
  Benjamin.
\newblock Variational-state quantum metrology.
\newblock {\em arXiv preprint arXiv:1908.08904}, 2019.

\bibitem{meyer2020variational}
Johannes~Jakob Meyer, Johannes Borregaard, and Jens Eisert.
\newblock A variational toolbox for quantum multi-parameter estimation.
\newblock {\em npj Quantum Information}, 7(1), Jun 2021.

\bibitem{PhysRevLett.118.150503}
Jun Li, Xiaodong Yang, Xinhua Peng, and Chang-Pu Sun.
\newblock Hybrid quantum-classical approach to quantum optimal control.
\newblock {\em Phys. Rev. Lett.}, 118:150503, Apr 2017.

\bibitem{yang2020probe}
Xiaodong Yang, Jayne Thompson, Ze~Wu, Mile Gu, Xinhua Peng, and Jiangfeng Du.
\newblock Probe optimization for quantum metrology via closed-loop learning
  control.
\newblock {\em npj Quantum Information}, 6(1):1--7, 2020.

\bibitem{schuld2019quantum}
Maria Schuld and Nathan Killoran.
\newblock Quantum machine learning in feature hilbert spaces.
\newblock {\em Physical review letters}, 122(4):040504, 2019.

\bibitem{kaubruegger2019variational}
Raphael Kaubruegger, Pietro Silvi, Christian Kokail, Rick van Bijnen, Ana~Maria
  Rey, Jun Ye, Adam~M Kaufman, and Peter Zoller.
\newblock Variational spin-squeezing algorithms on programmable quantum
  sensors.
\newblock {\em Physical Review Letters}, 123(26):260505, 2019.

\bibitem{kandala2017hardware}
Abhinav Kandala, Antonio Mezzacapo, Kristan Temme, Maika Takita, Markus Brink,
  Jerry~M Chow, and Jay~M Gambetta.
\newblock Hardware-efficient variational quantum eigensolver for small
  molecules and quantum magnets.
\newblock {\em Nature}, 549(7671):242--246, 2017.

\bibitem{anschuetz2019variational}
Eric Anschuetz, Jonathan Olson, Al{\'a}n Aspuru-Guzik, and Yudong Cao.
\newblock Variational quantum factoring.
\newblock In {\em International Workshop on Quantum Technology and Optimization
  Problems}, pages 74--85. Springer, 2019.

\bibitem{tannu19asplos}
Swamit~S. Tannu and Moinuddin~K. Qureshi.
\newblock Not all qubits are created equal: A case for variability-aware
  policies for nisq-era quantum computers.
\newblock In {\em Proceedings of the Twenty-Fourth International Conference on
  Architectural Support for Programming Languages and Operating Systems},
  ASPLOS '19, page 987–999, New York, NY, USA, 2019. Association for
  Computing Machinery.

\bibitem{murali19asplos}
Prakash Murali, Jonathan~M. Baker, Ali Javadi-Abhari, Frederic~T. Chong, and
  Margaret Martonosi.
\newblock Noise-adaptive compiler mappings for noisy intermediate-scale quantum
  computers.
\newblock In {\em Proceedings of the Twenty-Fourth International Conference on
  Architectural Support for Programming Languages and Operating Systems},
  ASPLOS '19, page 1015–1029, New York, NY, USA, 2019. Association for
  Computing Machinery.

\bibitem{murali19isca}
Prakash Murali, Norbert~Matthias Linke, Margaret Martonosi, Ali~Javadi Abhari,
  Nhung~Hong Nguyen, and Cinthia~Huerta Alderete.
\newblock Full-stack, real-system quantum computer studies: Architectural
  comparisons and design insights.
\newblock In {\em Proceedings of the 46th International Symposium on Computer
  Architecture}, ISCA '19, page 527–540, New York, NY, USA, 2019. Association
  for Computing Machinery.

\bibitem{poulami19micro}
Poulami Das, Swamit~S. Tannu, Prashant~J. Nair, and Moinuddin Qureshi.
\newblock A case for multi-programming quantum computers.
\newblock In {\em Proceedings of the 52nd Annual IEEE/ACM International
  Symposium on Microarchitecture}, MICRO '52, page 291–303, New York, NY,
  USA, 2019. Association for Computing Machinery.

\bibitem{murali20isca}
Prakash Murali, Dripto~M. Debroy, Kenneth~R. Brown, and Margaret Martonosi.
\newblock {\em Architecting Noisy Intermediate-Scale Trapped Ion Quantum
  Computers}, page 529–542.
\newblock IEEE Press, 2020.

\bibitem{braunstein1994statistical}
Samuel~L Braunstein and Carlton~M Caves.
\newblock Statistical distance and the geometry of quantum states.
\newblock {\em Physical Review Letters}, 72(22):3439, 1994.

\bibitem{fisher1925theory}
Ronald~Aylmer Fisher.
\newblock Theory of statistical estimation.
\newblock In {\em Mathematical Proceedings of the Cambridge Philosophical
  Society}, volume~22, pages 700--725. Cambridge University Press, 1925.

\bibitem{cramer1999mathematical}
Harald Cram{\'e}r.
\newblock {\em Mathematical methods of statistics}, volume~43.
\newblock Princeton university press, 1999.

\bibitem{greenberger1989going}
Daniel~M Greenberger, Michael~A Horne, and Anton Zeilinger.
\newblock Going beyond bell’s theorem.
\newblock In {\em Bell’s theorem, quantum theory and conceptions of the
  universe}, pages 69--72. Springer, 1989.

\bibitem{schulte2020ramsey}
Marius Schulte, Victor~J Mart{\'\i}nez-Lahuerta, Maja~S Scharnagl, and Klemens
  Hammerer.
\newblock Ramsey interferometry with generalized one-axis twisting echoes.
\newblock {\em Quantum}, 4:268, 2020.

\bibitem{powell1964efficient}
Michael~JD Powell.
\newblock An efficient method for finding the minimum of a function of several
  variables without calculating derivatives.
\newblock {\em The computer journal}, 7(2):155--162, 1964.

\bibitem{powell2007view}
Michael~JD Powell.
\newblock A view of algorithms for optimization without derivatives.
\newblock {\em Mathematics Today-Bulletin of the Institute of Mathematics and
  its Applications}, 43(5):170--174, 2007.

\bibitem{zhou2018achieving}
Sisi Zhou, Mengzhen Zhang, John Preskill, and Liang Jiang.
\newblock Achieving the heisenberg limit in quantum metrology using quantum
  error correction.
\newblock {\em Nature communications}, 9(1):1--11, 2018.

\bibitem{abraham1998introduction}
Raymond~John Abraham, Julie Fisher, and Philip Loftus.
\newblock {\em Introduction to NMR spectroscopy}, volume~2.
\newblock Wiley New York, 1998.

\bibitem{goldstein2011environment}
G~Goldstein, P~Cappellaro, JR~Maze, JS~Hodges, L~Jiang, Anders~S{\o}ndberg
  S{\o}rensen, and MD~Lukin.
\newblock Environment-assisted precision measurement.
\newblock {\em Physical Review Letters}, 106(14):140502, 2011.

\bibitem{bradley2019ten}
CE~Bradley, J~Randall, MH~Abobeih, RC~Berrevoets, MJ~Degen, MA~Bakker,
  M~Markham, DJ~Twitchen, and TH~Taminiau.
\newblock A ten-qubit solid-state spin register with quantum memory up to one
  minute.
\newblock {\em Physical Review X}, 9(3):031045, 2019.

\bibitem{bar2013solid}
Nir Bar-Gill, Linh~M Pham, Andrejs Jarmola, Dmitry Budker, and Ronald~L
  Walsworth.
\newblock Solid-state electronic spin coherence time approaching one second.
\newblock {\em Nature communications}, 4(1):1--6, 2013.

\bibitem{doherty2013nitrogen}
Marcus~W Doherty, Neil~B Manson, Paul Delaney, Fedor Jelezko, J{\"o}rg
  Wrachtrup, and Lloyd~CL Hollenberg.
\newblock The nitrogen-vacancy colour centre in diamond.
\newblock {\em Physics Reports}, 528(1):1--45, 2013.

\bibitem{boss2017quantum}
Jens~M Boss, KS~Cujia, Jonathan Zopes, and Christian~L Degen.
\newblock Quantum sensing with arbitrary frequency resolution.
\newblock {\em Science}, 356(6340):837--840, 2017.

\bibitem{barry2020sensitivity}
John~F Barry, Jennifer~M Schloss, Erik Bauch, Matthew~J Turner, Connor~A Hart,
  Linh~M Pham, and Ronald~L Walsworth.
\newblock Sensitivity optimization for nv-diamond magnetometry.
\newblock {\em Reviews of Modern Physics}, 92(1):015004, 2020.

\bibitem{dolde2011electric}
Florian Dolde, Helmut Fedder, Marcus~W Doherty, Tobias N{\"o}bauer, Florian
  Rempp, Gopalakrishnan Balasubramanian, Thomas Wolf, Friedemann Reinhard,
  Lloyd~CL Hollenberg, Fedor Jelezko, and J{\"o}rg Wrachtrup.
\newblock Electric-field sensing using single diamond spins.
\newblock {\em Nature Physics}, 7(6):459--463, 2011.

\bibitem{neumann2013high}
Philipp Neumann, Ingmar Jakobi, Florian Dolde, Christian Burk, Rolf Reuter,
  Gerald Waldherr, Jan Honert, Thomas Wolf, Andreas Brunner, Jeong~Hyun Shim,
  Dieter Suter, H.~Sumiya, Junichi Isoya, and J{\"o}rg Wrachtrup.
\newblock High-precision nanoscale temperature sensing using single defects in
  diamond.
\newblock {\em Nano letters}, 13(6):2738--2742, 2013.

\bibitem{kucsko2013nanometre}
Georg Kucsko, Peter~C Maurer, Norman~Ying Yao, MICHAEL Kubo, Hyun~Jong Noh,
  Po~Kam Lo, Hongkun Park, and Mikhail~D Lukin.
\newblock Nanometre-scale thermometry in a living cell.
\newblock {\em Nature}, 500(7460):54--58, 2013.

\bibitem{clevenson2015broadband}
Hannah Clevenson, Matthew~E Trusheim, Carson Teale, Tim Schr{\"o}der, Danielle
  Braje, and Dirk Englund.
\newblock Broadband magnetometry and temperature sensing with a light-trapping
  diamond waveguide.
\newblock {\em Nature Physics}, 11(5):393--397, 2015.

\bibitem{sushkov2014magnetic}
AO~Sushkov, I~Lovchinsky, N~Chisholm, Ronald~Lee Walsworth, Hongkun Park, and
  Mikhail~D Lukin.
\newblock Magnetic resonance detection of individual proton spins using quantum
  reporters.
\newblock {\em Physical review letters}, 113(19):197601, 2014.

\bibitem{schaffry2011proposed}
Marcus Schaffry, Erik~M Gauger, John~JL Morton, and Simon~C Benjamin.
\newblock Proposed spin amplification for magnetic sensors employing crystal
  defects.
\newblock {\em Physical review letters}, 107(20):207210, 2011.

\bibitem{glenn2018high}
David~R Glenn, Dominik~B Bucher, Junghyun Lee, Mikhail~D Lukin, Hongkun Park,
  and Ronald~L Walsworth.
\newblock High-resolution magnetic resonance spectroscopy using a solid-state
  spin sensor.
\newblock {\em Nature}, 555(7696):351--354, 2018.

\bibitem{sahni20045}
Varun Sahni.
\newblock 5 dark matter and dark energy.
\newblock In {\em The Physics of the Early Universe}, pages 141--179. Springer,
  2004.

\bibitem{preskill1983cosmology}
John Preskill, Mark~B Wise, and Frank Wilczek.
\newblock Cosmology of the invisible axion.
\newblock {\em Physics Letters B}, 120(1-3):127--132, 1983.

\bibitem{sikivie1983experimental}
Pierre Sikivie.
\newblock Experimental tests of the" invisible" axion.
\newblock {\em Physical Review Letters}, 51(16):1415, 1983.

\bibitem{pospelov2013detecting}
M~Pospelov, Szymon Pustelny, MP~Ledbetter, DF~Jackson Kimball, Wojciech Gawlik,
  and D~Budker.
\newblock Detecting domain walls of axionlike models using terrestrial
  experiments.
\newblock {\em Physical review letters}, 110(2):021803, 2013.

\bibitem{safronova2018search}
MS~Safronova, D~Budker, D~DeMille, Derek F~Jackson Kimball, A~Derevianko, and
  Charles~W Clark.
\newblock Search for new physics with atoms and molecules.
\newblock {\em Reviews of Modern Physics}, 90(2):025008, 2018.

\bibitem{ge2018distributed}
Wenchao Ge, Kurt Jacobs, Zachary Eldredge, Alexey~V Gorshkov, and Michael
  Foss-Feig.
\newblock Distributed quantum metrology with linear networks and separable
  inputs.
\newblock {\em Physical review letters}, 121(4):043604, 2018.

\bibitem{braverman2019near}
Boris Braverman, Akio Kawasaki, Edwin Pedrozo-Pe{\~n}afiel, Simone Colombo, Chi
  Shu, Zeyang Li, Enrique Mendez, Megan Yamoah, Leonardo Salvi, Daisuke
  Akamatsu, Yanhong Xiao, and Vladan Vuleti{\'c}.
\newblock Near-unitary spin squeezing in yb 171.
\newblock {\em Physical Review Letters}, 122(22):223203, 2019.

\bibitem{bothwell2019jila}
Tobias Bothwell, Dhruv Kedar, Eric Oelker, John~M Robinson, Sarah~L Bromley,
  Weston~L Tew, Jun Ye, and Colin~J Kennedy.
\newblock Jila sri optical lattice clock with uncertainty of $2\times10^{-18}$.
\newblock {\em Metrologia}, 56(6):065004, 2019.

\bibitem{nelder1965simplex}
John~A Nelder and Roger Mead.
\newblock A simplex method for function minimization.
\newblock {\em The computer journal}, 7(4):308--313, 1965.

\bibitem{fagaly2006superconducting}
RL~Fagaly.
\newblock Superconducting quantum interference device instruments and
  applications.
\newblock {\em Review of scientific instruments}, 77(10):101101, 2006.

\bibitem{penney2019survey}
Drew~D Penney and Lizhong Chen.
\newblock A survey of machine learning applied to computer architecture design.
\newblock {\em arXiv preprint arXiv:1909.12373}, 2019.

\end{thebibliography}

\end{document}